\documentclass[]{pasj01}

\begin{document}
\Received{}
\Accepted{}

\title{An application of the Ghosh \& Lamb model to the accretion powered X-ray pulsar X~Persei}


\author{Fumiaki \textsc{Yatabe}\altaffilmark{1,2}}
\altaffiltext{1}{High Energy Astrophysics Laboratory, RIKEN, 2-1 Hirosawa, Wako, Saitama 351-0198, Japan}
\email{fumiaki.yatabe@riken.jp}
\altaffiltext{2}{Department of Physics, College of Science, Rikkyo University, 3-34-1 Nishi-Ikebukuro, Toshima, Tokyo 171-8501, Japan}
\altaffiltext{3}{School of Dentistry at Matsudo, Nihon University, 2-870-1 Sakaecho-nishi, Matsudo, Chiba 101-8308, Japan}
\altaffiltext{4}{Department of Physics, Tokyo Institute of Technology, 2-12-1 Ookayama, Meguro-ku, Tokyo 152-8551, Japan}

\author{Kazuo \textsc{Makishima}\altaffilmark{1}}

\author{Tatehiro \textsc{Mihara}\altaffilmark{1}}

\author{Motoki \textsc{Nakajima}\altaffilmark{3}}

\author{Mutsumi \textsc{Sugizaki}\altaffilmark{4}}

\author{Shunji \textsc{Kitamoto}\altaffilmark{2}}

\author{Yuki \textsc{Yoshida}\altaffilmark{2}}

\author{Toshihiro \textsc{Takagi}\altaffilmark{1}}




\maketitle

\begin{abstract}
The accretion-induced pulse-period changes of the Be/X-ray binary pulsar X~Persei were investigated over a period of 1996 January to 2017 September.
This study utilized the monitoring data acquired with the RXTE/ASM in 1.5--12 keV and MAXI/GSC in 2--20 keV. 
The source intensity changed by a factor of 5--6 over this period.
The pulsar was spinning down for 1996-2003, and has been spinning up since 2003, as already reported.
The spin up/down rate and the 3--12 keV flux, determined every 250 d, showed a clear negative correlation, which can be successfully explained by the accretion torque model proposed by \citet{GL79}. 
When the mass, radius and distance of the neutron star are allowed to vary over a range of 1.0--2.4 solar masses, 9.5--15 km, and 0.77--0.85 kpc, respectively, the magnetic field strength of $B=(4-25) \times10^{13}\ \rm G$ gave the best fits to the observation. 
In contrast, the observed results cannot be explained by the values of $B\sim10^{12}\ \rm G$ previously suggested for X~Persei, as long as the mass, radius, and distance are required to take reasonable values.
Assuming a distance of $0.81\pm0.04$ kpc as indicated by optical astrometry, the mass of the neutron star is estimated as $M=2.03\pm0.17$ solar masses.
\end{abstract}

\section{Introduction}
The magnetic fields of binary X-ray pulsars have been directly measured utilizing cyclotron resonance scattering features (hereafter CRSF) in their X-ray spectra (e.g., \cite{Makishima2016}).
The magnetic field strength $B$ measured in this way is distributed in the range of $B/(1+{z_{\rm g}})= (1-7)\times 10^{12}$ G (\cite{Yamamoto2014} and references therein), where $z_{\rm g} \simeq 0.24$ is the gravitational redshift at the neutron-star (hereafter NS) surface.
No accreting pulsars with $B\geq10^{13}$ G have yet been discovered.
However, this could be due to a selection effect, that the CRSF becomes difficult to detect in $\geq$ 100 keV (i.e, for $B\geq10^{13}$ G) where the observational sensitivity becomes progressively lower.
In order to search for such accreting NSs with higher magnetic fields, we need to invoke other methods such as pulse timing analysis or modeling of continuum spectra \citep{Sasano2015}.

Using the MAXI Gas Slit Camera (GSC) data and some past measurements, \citet{Takagi2016} investigated long-term relations between the pulse period derivative $\dot{P}$ and the fluxes of the binary X-ray pulsar 4U~1626-67.
They adopted the accretion torque theory proposed by \citet{GL79} (hereafter GL79), which describes $\dot{P}$ as a function of the luminosity $L$, the pulse period $P$, the mass $M$, the radius $R$, and the surface magnetic field $B$ of the NS.
Then, employing $B/(1+{z_{\rm g}})=3.2\times 10^{12}$ G measured with a CRSF \citep{Orland1998}, the GL79 model was confirmed to accurately explain the observed relation between $\dot{P}$ and $L$ of 4U 1626-67, including both the spin-up and spin-down phases.
This means that accurate measurements of $\dot{P}$ over a sufficiently wide range of $L$ of a binary X-ray pulsar would conversely allow us to estimate its $B$.

X~Persei (4U 0352+309) is a high-mass X-ray binary system, consisting of a Be-type star and a magnetized NS. 
Its pulse period of $P\sim835\ \rm s$, which is relatively longer than those of other pulsars, was independently discovered by Ariel 5 and Copernicus \citep{White1976}.
The distance to the source is estimated as $D=0.81\pm0.04$ kpc by its optical parallax derived from the GAIA Astrometry \footnote{https://gea.esac.esa.int/archive/}.
With X-ray observations, \citet{Delgado2001} determined the orbital period of X~Persei as $P_{\rm orb}=250.3 \pm 0.6$ d, and derived an eccentricity of $e\sim$0.11 which is considerably smaller than those of typical Be/X-ray binaries, $e\geq$0.3.

X~Persei was spinning up at a rate of $\dot{P}\sim -1.5 \times 10^{-4}\ \rm yr^{-1}$ until 1978, when it turned into a phase of spin down at a rate of $\dot{P}\sim 1.3 \times 10^{-4}\ \rm yr^{-1}$ \citep{Delgado2001}.
In 2002, the pulsar returned to the spin-up phase, and has since been spinning up until today.
This spin up/down alternation implies that the source is close to a torque equilibrium, and hence the long $P$ suggests a strong magnetic field, because we then expect $B\sim L^{\frac{1}{2}}P^{\frac{6}{7}}$ \citep{Makishima2016}

In the 2--20 keV range, the spectrum of X~Persei can be fitted by a powerlaw model with an exponential cutoff \citep{DiSalvo1998}, which is typical of accreting X-ray pulsars.
However, in the 20--100 keV band, the spectrum of X~Persei is much flatter and harder than those of other X-ray pulsars, and lacks steep cutoff \citep{Sasano2015}. 
\citet{Sasano2015} empirically quantified this shape of the spectrum in comparison with those of other pulsars (mostly with $B$ measured), and obtained an estimate of of $B\sim10^{13}\ \rm G$. 
Although \citet{Coburn2001} noticed a spectral dip at $\sim 29$ keV in an RXTE/PCA spectrum, and regarded it as a CRSF corresponding to $B/(1+{z_{\rm g}})=2.5 \times 10^{12}\ \rm G$, the feature is too shallow and broad for that interpretation, and can be explained away by a combination of two continuum components without a local feature \citep{Doroshenko2012, Sasano2015}. Thus, the strong-field property of X~Persei is suggested also by its spectral properties.

In the present paper, we report the long-term continuous observations of the fluxes and $P$ of X~Persei by RXTE and MAXI.
Applying the GL79 model to the derived $\dot{P}-L$ relation, we then estimate $B$ and $M$ of the NS in this binary.


\section{Observation}

\begin{figure*}
 \begin{center}
  \includegraphics[width=0.7\textwidth]{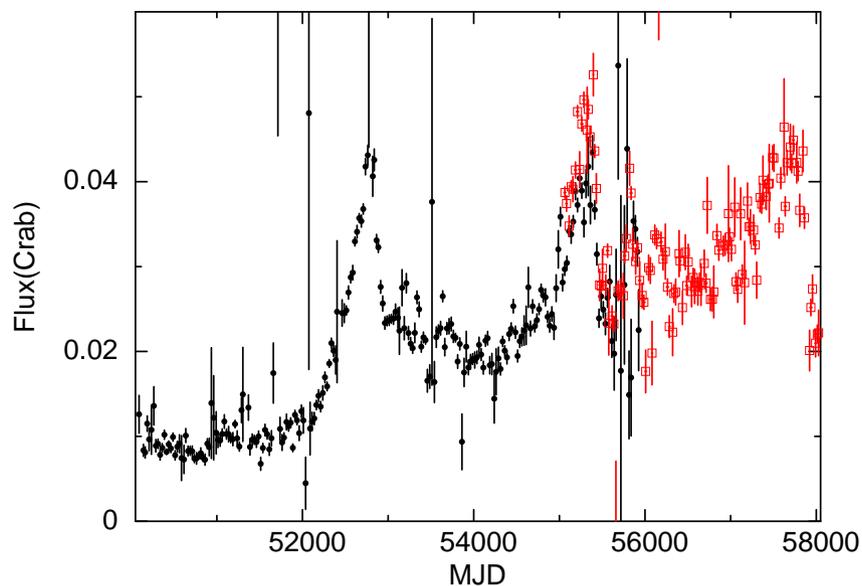}
 \end{center}
\caption{Long term (1996--2017) light curves of X~Persei in 25-d bin, expressed in Crab units. Black filled circles and red open squares are data with the RXTE/ASM (1.5--12 keV) and the MAXI/GSC (2--20 keV), respectively.}
\label{longlc}
\end{figure*}

\subsection{MAXI/GSC}
Since August 2009, MAXI \citep{Matsuoka2009} on board the International Space Station (ISS) has been continuously scanning the X-ray sky every 92 min of the ISS orbital period.
We derived the data of X~Persei from the MAXI/GSC \citep{Mihara2011, Sugizaki2011}, using On-Demand System \footnote{https://maxi.riken.jp/mxondem/} provided by the MAXI team.
This data-analysis scheme enables us to extract images, light curves, and energy spectra of a given source by specifying its coordinate and observation times.
The on-source events and background events, both in 2--20 keV were derived using standard regions.
Every 92 min, we thus obtained a background-subtracted data set (an energy spectrum), with an integration time of $\sim 60\ \rm s$ corresponding to a single scan transit of the source.
Since this is sufficiently shorter than compared to the pulse period of X~Persei ($\sim835\ \rm s$), the data obtained with a single-scan is just a snap shot at a certain pulse phase.
These data sets have been combined into a 2--20 keV light curve from MJD 55054 to MJD 58049 and 12 energy spectra each summed over the 250-d orbital period.
The 250-d averaged spectra have statistics high enough to yield accurate fluxes.

\subsection{RXTE/ASM}
The All-Sky monitor (ASM; \cite{Levine1996}) on board the RXTE satellite \citep{Bradt1993} continuously monitored the whole X-ray sky from MJD 50087 to MJD 55927.
For this entire period, we retrieved the RXTE/ASM dwell-time light curves of X~Persei in four energy ranges, namely, 1.5--12.0 keV, 1.5--3.0 keV, 3.0--5.0 keV, and 5.0--12.0 keV\footnote{https://xte.mit.edu/ASM\_lc.html/}.
These light curves were used to obtain $P$ and the fluxes.

\subsection{RXTE/PCA}
In order to cross check the validity of the fluxes of RXTE/ASM and MAXI/GSC, RXTE/PCA data were also used.
Since the PCA data are generally too short and sparse for the accurate determination of $P$ and calculation of $\dot{P}$, these data were not used for pulse period analysis.
Through 16 years RXTE mission, total 156 pointing observations were performed on this source by the Proportional Counter Array (PCA; \cite{Jahoda2006}). 
Those PCA observations were mede in the period from MJD 50161 to MJD 52687.
In this duration, one of the PCA unit, PCU 2, had longer exposure times than the others.
Thus, we used the data taken by PCU 2 only.
An inspection of the operation status of PCU 2, indicated that some of the data sets were acquired under bad conditions (e.g., large offset angles). 
We discarded those data sets, and utilized 136 pointing observations (total exposure time of 640 ksec) to extract the spectra.

\subsection{Long-term light curves}
To investigate intensity variations of X~Persei, we have binned all the RXTE/ASM and MAXI/GSC data into 25-d bin light curves.
The derived long-term light curves from 1996 to 2017 are shown in figure \ref{longlc}, where the background-subtracted count rate is normalized to that of the Crab Nebula.
Thus, X~Persei showed moderate X-ray intensity variations up to a factor of 5-6.
The intensity is not apparently modulated at the orbital period, presumably because of the small eccentricity.
Instead, in 2003, 2010, and 2017, the intensity increased to 50 mCrab and dropped to 25 mCrab in one orbital period, suggesting a 7 years super-orbital periodicity.

\section{Analysis and results}

In order to calculate $\dot{P}$, it is necessary to first determine $P$ accurately, with a reasonably dense sampling.
For this purpose, the times of individual light-curve bins were converted to those to be measured at the solar center; this heliocentric correction, instead of the more complex barycentric correction, is sufficient for the present purpose because of the long pulse period.
Then, we performed time corrections for the binary orbital motion of X~Persei, assuming the circular orbit parameters ; $P_{\rm orb}=249.9$ d, the epoch of $90^\circ$ mean orbital longitude : $T_{\pi/2} = 51215.5$ (MJD), and the semi-major axis $a_{\rm x} \mathrm{\,sin}\,i =454$ lt-s, from Table 2 in \citet{Delgado2001}.
The small eccentricity (e$\sim$0.11) affects the value of $P$ at most by $\pm0.010$ s in 70-d analysis (Section 3.1) and $\pm0.001$ s in 250-d analysis (Section 3.2) ; these are smaller than typical errors of the $P$ measurements.

\subsection{Refinement of the orbital period}

\begin{figure*}
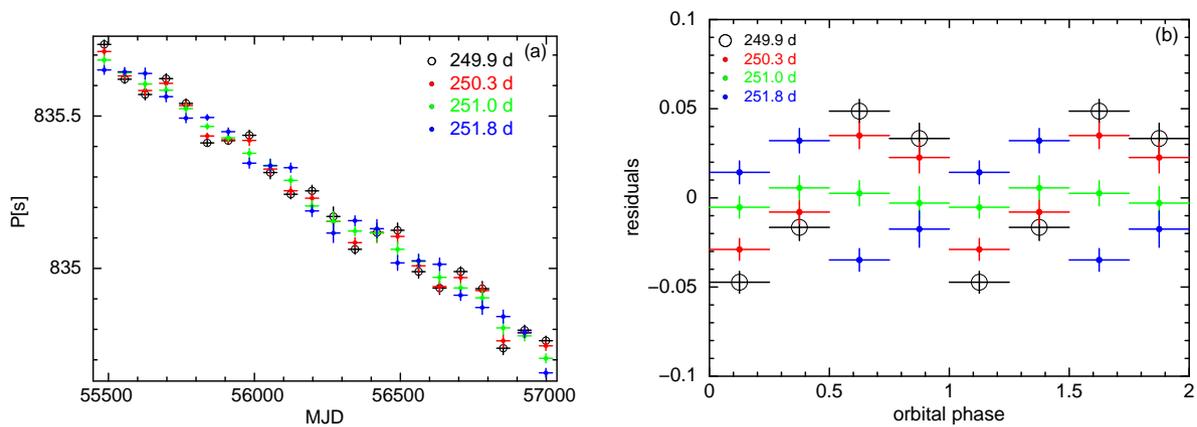

 \begin{center}
  \includegraphics[width=8cm]{p_70days_representative_4point_90.ps}
    \includegraphics[width=8cm]{folded_residuals_representative_4point_90.ps} 
 \end{center}
\caption{(a)Pulse periods of X~Persei determined every 70 d with the MAXI/GSC. Colors specify different values of $P_{\rm orb}$ employed to correct the pulse arrival times for the orbital motion of the NS.
(b)Pulse-period residuals, obtained from panel (a) by subtracting a linear trend and folding at the assumed $P_{\rm orb}$.}
\label{pulse70d}
\end{figure*}

\begin{figure}
 \begin{center}
  \includegraphics[width=0.5\textwidth]{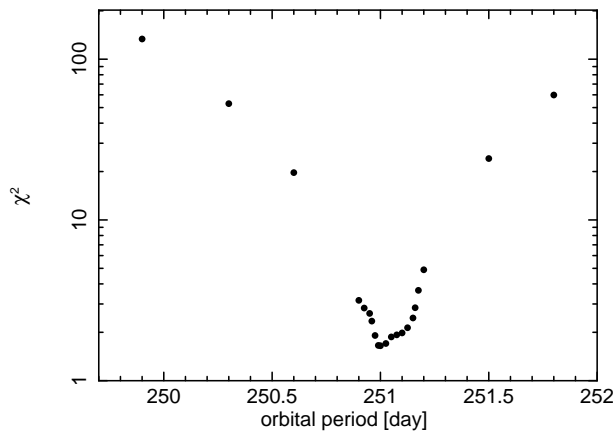}
 \end{center}
\caption{The values of $\chi^2$ from a constant fit to the residual profiles such as shown in figure \ref{pulse70d}, presented as a function of the assumed $P_{\rm orb}$}
\label{orbchi}
\end{figure}

As a preliminary attempt to confirm the arrival-time corrections, we determined the pulse period every 70 d for MJD55450--57020, when the luminosity stayed relatively constant.
The epoch folding method described in section 3.2 was employed, assuming $\dot{P}=-7\times10^{-9}\ \rm s\ s^{-1}$ which is the average value of $\dot{P}$ during this time period.
The result is shown with black points in figure \ref{pulse70d}.
Although the values of $P$ confirm the spin-up trend in general, the data points show wiggling behavior around the trend, with a typical period of 4 data points, or $\sim280$ d.
Since the flux in figure \ref{longlc} is not modulated at this period, the effect is unlikely to be caused by accretion-torque changes.
Instead, it could be due to slight discordance of the orbital phase, which in turn arose from an accumulation of the error of $P_{\rm orb}$ for 12--16 years since the observation by \citet{Delgado2001}.
We searched for a better orbital period as follows assuming that $T_{0} = 51215.5 $ (MJD) and $a_{\rm x} \mathrm{\,sin}\,i =454$ lt-s are correct.
First we repeated this analysis by changing the trial value of $P_{\rm orb}$ from 249.9 d to 251.8 d with a step of 0.3 d or finer.
Then, from the pulse-period history obtained in this way, we subtracted a trend line represented by $\dot{P}=-7\times10^{-9}\ \rm s\ s^{-1}$.
Finally, the residual pulse period was folded at the employed $P_{\rm orb}$.
Black points in figure \ref{pulse70d} show the folded residuals assuming, for example, $P_{\rm orb}=249.9$ d, where large residuals make a constant-line fit unacceptable with $\chi^2=134$ (for d.o.f.= 3).
The values of $\chi^2$ derived from this $P_{\rm orb}$ scan are plotted in figure \ref{orbchi}, and the orbital profile of the residuals for four representative cases (including the one with $P_{\rm orb}=249.9$ d) are presented in figure \ref{pulse70d}.
Requiring that the residual pulse-period modulation at the assumed $P_{\rm orb}$ should be minimized, the best orbital period was estimated as $P_{\rm orb}=251.0^{+0.2}_{-0.1}$ d ($1 \sigma$ error) which is still not away from the $2 \sigma$ error range given by \cite{Delgado2001}.
We hereafter use this value for the binary orbital motion corrections.

\subsection{Pulse timing analysis}

\subsubsection{Measurements of $P$}
Now that $P_{\rm orb}$ has successfully been updated, we employed the best orbital period in the correction of the binary motion.
Then, we analyze the entire RXTE/ASM and MAXI/GSC data of X~Persei, via the standard epoch folding method \citep{Leahy1983} to determine $P$ of X~Persei every 250 d.
This time period, which is longer than employed in Section 3.1 and is close to $P_{\rm orb}$, has been chosen to accurately estimate $\dot{P}$, which is important in the GL79 formula.
The employed energy band is 1.5--12 keV for the RXTE/ASM and 2--20 keV for MAXI/GSC. 
The number of bins of the folded pulse profile was chosen as 16, because one MAXI data point is an accumulation over about 60 s.
We assumed a constant but free $\dot{P}$ during each 250-d time period, and searched for a pair ($P$, $\dot{P}$) that maximizes $\chi^2$ of the folded 16-bin pulse profile against the constant-intensity hypothesis.

An example of $\chi^2$ map on the ($P$, $\dot{P}$) is shown in figure \ref{cont}(a), and the corresponding folded pulse profile using ($P$, $\dot{P}$) at the maximum $\chi^2$ is shown in figure \ref{cont}(b).
The pulse profile has a sinusoidal shape, not only in this particular case, but also in other intervals.

The derived values of $P$ are shown in figure \ref{fp}(b), together with the 250-d averaged 3--12 keV flux to be described in the next section.
The 1$\sigma$ errors of $P$ were estimated with the Monte-Carlo method of \citet{Leahy1987}.
As \citet{Lutovinov2012} and \citet{Acuner2014} reported, the pulsar changed from a spin-down phase to a spin-up phase around MJD52000, when the X-ray flux started increasing.
Since then, X~Persei has been in the spin-up phase at least until MJD58049.

\subsubsection{Calculations of $\dot{P}$}
\begin{figure*}
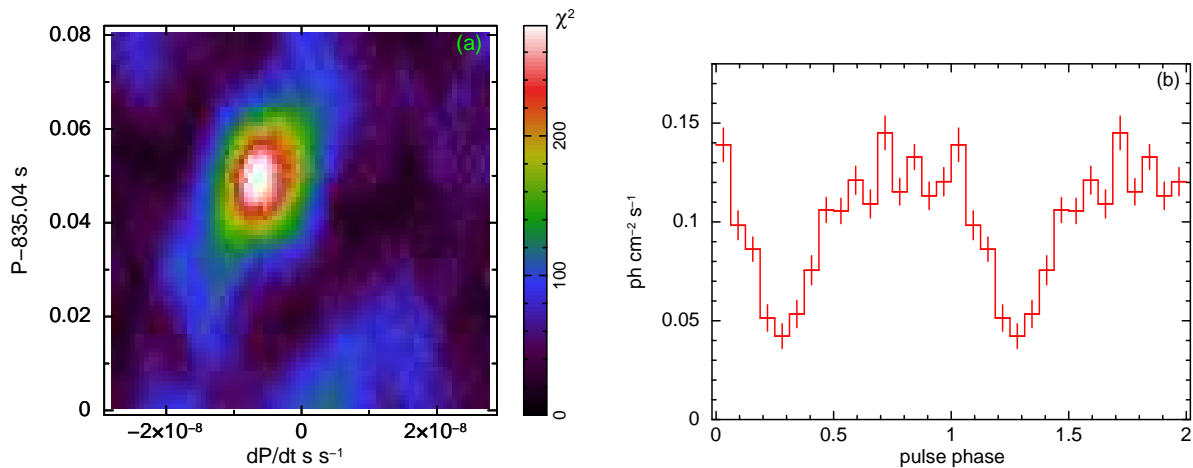

 \begin{center}
  \includegraphics[width=8cm]{cont_2-20keV_56425_2_1_90.ps} 
   \includegraphics[width=8cm]{pulse_profile_2-20keV_56300-56549_90.ps} 
 \end{center}
\caption{(a) A color map of the 2-20 keV pulse significance in terms of $\chi^2$, for MJD56300--56549, shown on a $P-\dot{P}$ plane. The best-fit values are $P=835.090 \pm 0.002\ \rm s$ and $\dot{P}=-(6.3\pm0.7)\times10^{-9}\ \rm s\ s^{-1}$.
(b) A background-subtracted 2--20 keV pulse profile of X~Persei in MJD56300--56549, folded with the ${\chi^2}$-maximum parameters.}
\label{cont}
\end{figure*}

Although $\dot{P}$ was thus estimated every 250 d, the calculation of $\dot{P}$ can also be done by taking the difference between adjacent two points of $P$.
Let us denote $\dot{P}$ from the $P$--$\dot{P}$ plane (e.g., figure \ref{cont}), as $\dot{P}_{\chi^2}$, and $\dot{P}$ from the difference between adjacent $P$ measurements as $\dot{P}_{\rm \ diff}$.
Then, as shown in figure \ref{fp}(c), $\dot{P}_{\rm \ diff}$ is consistent with $\dot{P}_{\chi^2}$, and has smaller uncertainty.
Therefore we employ $\dot{P}_{\rm \ diff}$ hereafter.

\subsection{Calculation of the energy flux}
The energy flux of X~Persei at individual epochs can be derived from the MAXI/GSC and the RXTE/ASM data.
In addition, a limited amount of data from the RXTE/PCA are utilized.
In order to minimize systematic differences among the three instruments, we decided to use their common energy band, 3--12 keV.
Below, we describe how the fluxes were derived with the three instruments.

\subsubsection{The MAXI/GSC data}
The 2--20 keV energy spectra of X~Persei with the MAXI/GSC were accumulated over the same 250-d time periods as section 3.2, and fitted individually by a power-law with an exponential cutoff.
The intersteller absorption was fixed to $N_{\rm H}=3.4\times10^{21}\,\rm cm^{-2}$ derived from an XMM-Newton observation \citep{Palombara2007}, because it cannot be determined independently with the MAXI/GSC.
The spectral fits were acceptable in all the time periods, with the reduced chi-square of $\chi_{\rm \nu}^2 \sim 1$.
We then calculated the unabsorbed 3--12 keV fluxes from the best-fit models.
For example, the best-fit model for MJD 56300--56549 has a photon index of $\Gamma=0.67\pm 0.13$, a cutoff energy of $5.57\pm0.76\ \rm keV$, and a normalization of $(8.4\pm0.7)\times 10^{-2}\ \rm $ with $\chi^2=119$ for 105 d.o.f.

By analyzing all the MAXI/GSC spectrum in the same way, the 2--20 keV fluxes were obtained as $(9-14)\times10^{-10}$ erg cm$^{-2}$ s$^{-1}$, and those in 3--12 keV as $(6-10)\times10^{-10}$ erg cm$^{-2}$ s$^{-1}$.
The latter results are presented with red squares in figure \ref{fp}(a), in comparison with the $\dot{P}$ determinations made in subsection 3.3. (A correction factor of 0.78 was multiplied in order to match the data to the RXTE/ASM data. See Section 3.4.2.)
In figure \ref{fp}(a), we can reconfirm the intensity variations seen in figure \ref{longlc}.

\subsubsection{The RXTE/ASM data}
By adding the dwell time light curves in 3--5 keV and 5--12 keV, the 3--12 keV count rates averaged over the individual 250-d time periods were obtained. 
We converted the 3--12 keV count rates into 3--12 keV fluxes using WebPIMMS system\footnote{https://heasarc.gsfc.nasa.gov/cgi-bin/Tools/w3pimms/w3pimms.pl}.
Since this conversion requires the knowledge of spectrum shape of the source, we followed the results of the MAXI spectroscopy, and assumed a power-law spectrum with $\Gamma = 1.7$ and $N_{\rm H}=3.4\times10^{21}\,\rm cm^{-2}$ \citep{Palombara2007} in all the 250-d time periods.
Although the MAXI spectra covers only the spin-up phase after 2009, our assumption on the spectral shape is justified by \citet{Lutovinov2012}, who analyzed the 2--20 keV RXTE/PCA data of X~Persei in both the spin-down and spin-up phases, and found no significant difference in the spectral shape.

\begin{figure*}
 \begin{center}
 \includegraphics[width=0.6\textwidth]{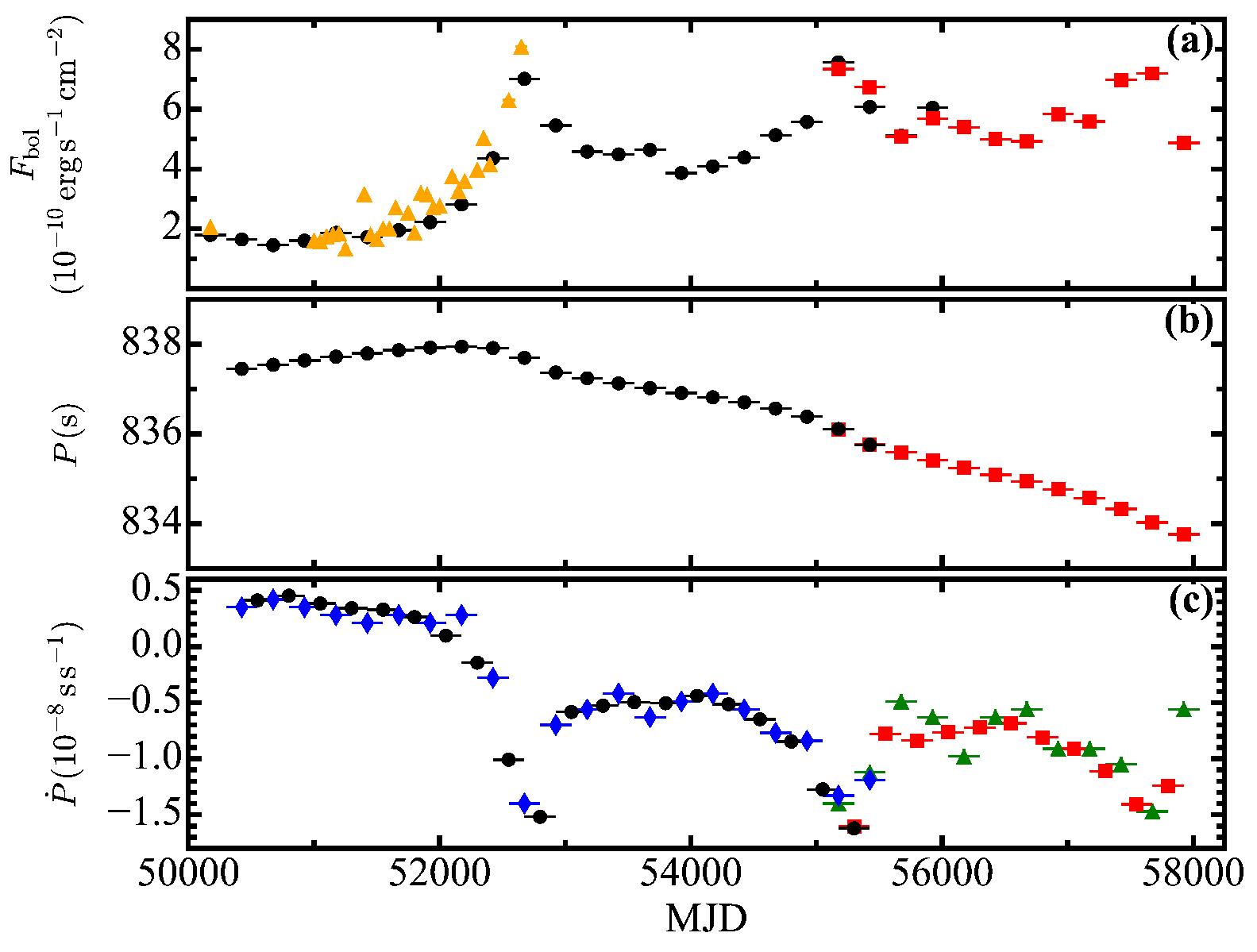}
 \end{center}
\caption{(a) The 3--12 keV fluxes of X~Persei obtained with the RXTE/ASM (black circles), and the MAXI/GSC (red squares), and the RXTE/PCA (yellow triangles). The MAXI/GSC fluxes have been converted as described in text.
(b) The intrinsic pulse period of X~Persei, after the heliocentric and binary orbit corrections. In both panel(a) and panel(b), one data point covers a 250-d time interval.
(c) A comparison between $\dot{P}_{\chi^2}$ (blue diamonds and green triangles) and $\dot{P}_{\rm diff}$ (black and red).}
\label{fp}
\end{figure*}

The unabsorbed fluxes thus estimated with the RXTE/ASM data are shown in figure \ref{fp}(a) by black filled circles.
Over the time range when the MAXI and RXTE/ASM results overlap, we found that the MAXI fluxes are systematically higher, by $\sim25$\%, than those of the RXTE/ASM.
Since the RXTE/ASM fluxes are consistent with BeppoSAX observation \citep{DiSalvo1998} and the RXTE/PCA measurements described in the next section, the MAXI/GSC fluxes have been corrected to match the RXTE/ASM measurements by multiplying a factor 0.78.
The investigation of this difference is beyond the scope of this work.

\subsubsection{The RXTE/PCA data}
First, we investigated the individual spectra obtained by the 136 pointing observations.
  Although some of them have short exposure times ($\leq1$ ksec), these data were confirmed to have sufficient quality to analyze the 3--12 keV X-ray spectra.
  The PCU2 spectra were fitted by a power-law model with an exponential cut-off, incorporating a low-energy absorption factor with $N_{\rm H}=3.4\times10^{21}\,\rm cm^{-2}$ (again fixed).
  The model successfully describe all the PCU2 spectra.
  Next, we calculated unabsorbed 3--12 keV fluxes of all these observations, and averaged them typically every $\sim 50$ d.
  The results are shown in figure \ref{fp}(a) with yellow triangles.
  Thus, the PCA and ASM results are consistent with each other, justifying the correction of the MAXI/GSC fluxes.
  As already noted in section 2.3, we do not utilize the PCA data in the pulse-timing studies.

\subsubsection{Estimation of the bolometric flux}
Since the luminosity used in the GL79 formalism is the bolometric value $L_{\rm bol}$, we converted the 3--12 keV fluxes to the bolometric ones, $F_{\rm bol}$.
For this purpose, we employed the best-fit model to the 0.1--200 keV spectrum obtained with BeppoSAX in 1996 September (spin-down phase), and that in the 1--100 keV band obtained with Suzaku in 2012 September (spin-up phase), derived by \citet{DiSalvo1998} and \citet{Sasano2015}, respectively.
The conversion factor from the 3--12 keV to 0.1--200 keV fluxes was calculated as 2.605 in BeppoSAX and 2.611 in Suzaku, respectively.
Since the two factors agree very well within $\leq0.1$\%, in agreement with \citet{Lutovinov2012}, we used a factor of 2.61 for both the spin-down and spin-up phases, and multiplied it to all the data points in figure \ref{fp}(a).

\subsubsection{The correlation between $\dot{P}$ and $F_{\rm bol}$}
The derived $\dot{P}$ and $F_{\rm bol}$ values of X~Persei, spanning 22 years, are shown in figure \ref{GL}.
Thus, we find a clear negative correlation between $\dot{P}$ and $F_{\rm bol}$ of X~Persei.
Although a few data points are outlying, $\dot{P}$ basically behaves as a single-valued function of $F_{\rm bol}$ as the latter varies by a factor of $\sim 5$.
After renormalizing the MAXI/GSC fluxes (section 3.3.2), the data points from the two instruments are generally consistent with one another.
Furthermore, a single correlation appears to persist through the spin-up and spin-down phases, without exhibiting any drastic changes towards lower fluxes.
Therefore, the system is considered to be still free from so-called propeller effect, which might set in at very low accretion rates.
Thus, the overall source behavior in figure \ref{GL} is very similar to the case of 4U~1626--67 studied by \citet{Takagi2016}, wherein the GL79 model has given a very successful explanation.

\section{Application of GL79 model}
\subsection{The comparison between the observational results and GL79}
\begin{figure}
 \begin{center}
  \includegraphics[width=0.5\textwidth]{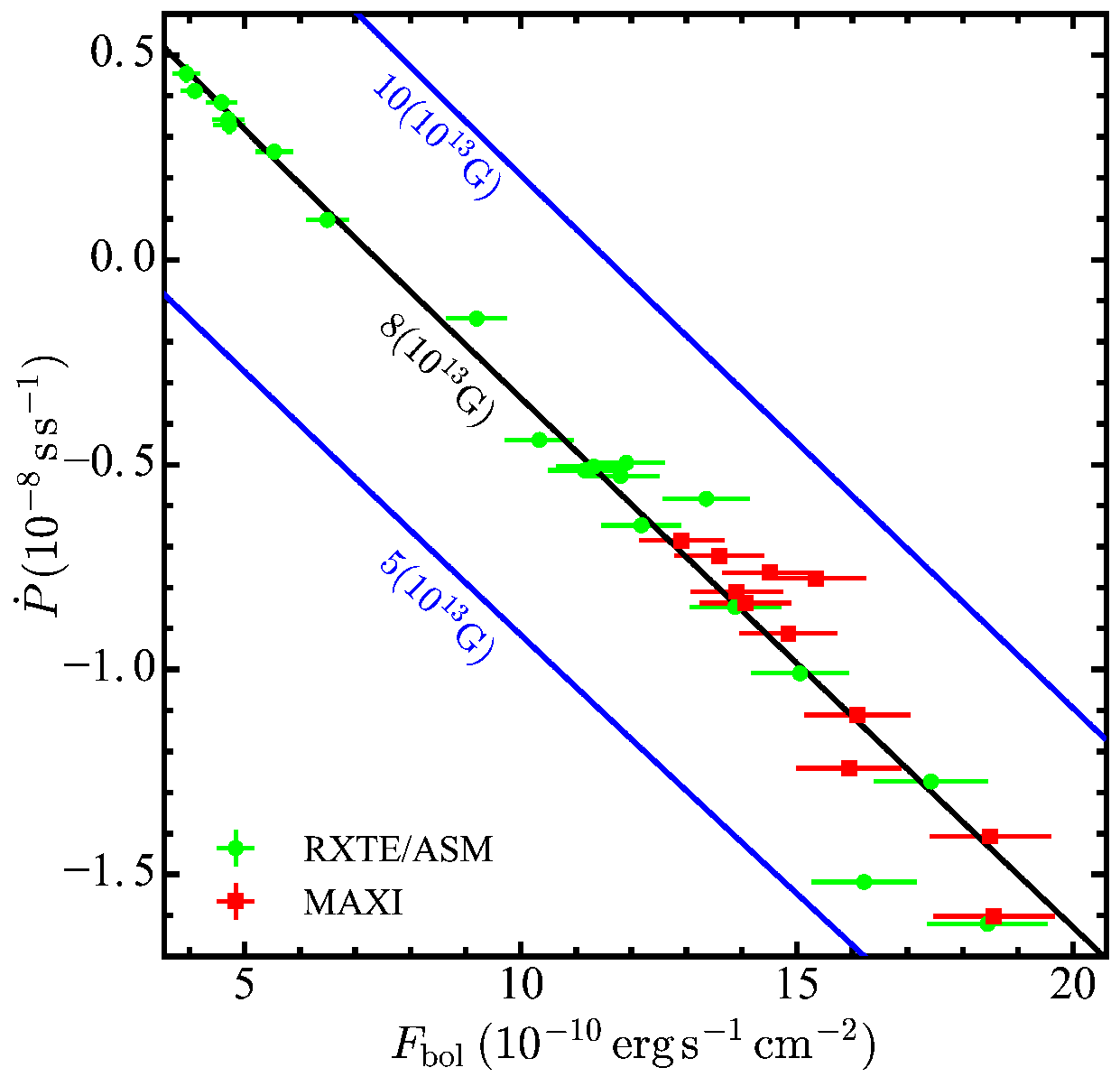}
 \end{center}
\caption{The $\dot{P}$ and $F_{\rm bol}$ measurements with the RXTE/ASM (green circles), and MAXI/GSC (red squares), compared with the GL79 predictions. The black line shows the best-fit GL79 relation. It can be represented by multiple sets of parameters (figure \ref{massradi}), including a particular case of $M=2.05M_{\odot}$, $R=12.9$ km, $D=0.81$ kpc, and $B=8\times 10^{13}$ G. The two blue lines show predictions when $B$ is changed, with the other parameters kept unchanged.}
\label{GL}
\end{figure}

The final step of our analysis is to fit the $\dot{P}$--$F_{\rm bol}$ relation in figure \ref{GL} with the GL79 model, to constrain the parameters of the NS in X~Persei; $M$, $R$ and $B$.
We use the same procedure as described in Appendix 1 in \citet{Takagi2016}, as briefly reviewed below.

According to GL79, $\dot{P}$ is described as
\begin{equation}
\dot{P}=-5.0 \times 10^{-5}\ \mu_{30}^{\frac{2}{7}}\ n(\omega_{\rm s})\ R_{6}^{\frac{6}{7}}\ \left(\frac{M}{M_{\odot}}\right)^{-\frac{3}{7}} I_{45}^{-1}\ P^{2}\ L_{37}^{\frac{6}{7}}\ {\rm \ s\ yr}^{-1},
\label{eqgl}
\end{equation}
where $\mu_{30}$, $R_{6}$, $M_{\odot}$, $I_{45}$, $P$, and $L_{37}$ are the magnetic dipole moment in units of $10^{30}{\rm\ G\ cm^{3}}$, $R$ in unit of ${10^{6}\rm\ cm}$, the solar mass, the moment of inertia in unit of ${10^{45}\rm\ g\ cm^2}$, the pulse period in unit of s, and the luminosity in units of ${10^{37}\rm\ erg\ s^{-1}}$.
The factor $n(\omega_{\rm s})$ describes corrections for the accretion torque exerted onto the NS from the accreting matter via disk.
It is represented as a function of the fastness parameter, $\omega_{\rm s}$, which means the ratio of the angular velocity of the NS rotation to that of the Kepler rotation at the inner disk radius.
In the GL79 model, $n(\omega_{\rm s})$ is approximately given by

\begin{equation}
n(\omega_{\rm s})\simeq1.39\{1-\omega_{\rm s}[4.03(1-\omega_{\rm s})^{0.173}-0.878]\}(1-\omega_{\rm s})^{-1},
\label{nomega}
\end{equation}
with
\begin{equation}
\omega_{\rm s}\sim1.35\ \mu_{30}^{\frac{6}{7}}\ R_{6}^{-\frac{3}{7}}\left(\frac{M}{M_{\odot}}\right)^{-\frac{2}{7}}P^{-1}\ L_{37}^{-\frac{3}{7}}
\label{omega}
\end{equation}
in the accretion regime of $0\leq \omega_{\rm s} \leq0.9$.
The spin-up and spin-down regimes corresponds to $\omega_{\rm s} \leq0.349$ and $\omega_{\rm s}\geq0.349$, respectively, with $\omega_{\rm s}=0.349$ describing a torque equilibrium.

We used a approximation for $I_{45}$ proposed by \citet{Lat2005} as
\[
I_{45} \simeq (0.474\pm0.016) \left(\frac{M}{M_{\odot}}\right)R_{6}^2
\]
\begin{equation}
\ \ \ \ \ \ \ \ \ \ \times \left[1+0.42\left(\frac{M}{M_{\odot}}\right)R_{6}^{-1}+0.009\left(\frac{M}{M_{\odot}}\right)^{4} R_{6}^{-4}\right].
\end{equation}
Considering the relativistic effects, $\mu$ is described in cgs units \citep{Was1983} as
\begin{equation}
\mu = \frac{1}{2} {BR^3}\ \frac{X^3}{3}\left[-\rm ln(1-X)-X-\frac{X^2}{2}\right]^{-1}
\end{equation}
where $X\equiv \left({R}/{R_{\rm s}}\right)^{-1}$, and $R_{\rm s}=2GM/c^2$ is the Schwarzschild radius.
If $R\gg R_{\rm s}$, this formula can be expanded
\begin{equation}
\mu \simeq \frac{1}{2} {BR^3}\left[1+\frac{3}{4}X+\frac{3}{5}X^{2}+\cdots \right]^{-1}.
\end{equation}
Finally, using the known $D$ to X~Persei, $L$ is expressed as
\begin{equation}
L=4 \pi D^2 F_{\rm bol}
\end{equation}

By substituting equations (2) through (7) to equation (1), we can calculate a theoretical $\dot{P}$ vs $L_{37}$ relation, to fit the observed data points in figure \ref{GL}.
However, the prediction is subject to uncertainties of the parameters involved, namely, $M$, $R$, $D$, and $B$.
Because $B$ is least constrained among them, and of our prime interest, we restricted ranges of $M$, $R$ and $D$ to 1.0--2.4$M_{\odot}$, 9.5--15 km, (e.g., \cite{Ozel2016, Bauswein2017}) and $0.81\pm 0.04$ kpc (Section 1), respectively, and aimed to estimate $B$.
Since the correlation is almost linear which has two free parameters, we specified trial values of ($B$, $D$), and optimized ($M$, $R$) to minimize the $\chi^2$.

An example of the best-fit GL79 prediction is shown with a black line in figure \ref{GL}, where $B=8\times 10^{13}$ G was assumed and the best-fit parameters was obtained as ($M$, $R$, $D$) = ($2.05M_{\odot}$, 12.9 km, 0.81 kpc).
By introducing a 5.9\% systematic error to $F_{\rm bol}$, the data points have been explained successfully ($\chi^2 =29$ d.o.f = 29), including both the spin-up and spin-down phases, since $n(\omega_{\rm s})$ can take both positive and negative values.
However, due to the heavy model degeneracy, the parameters cannot be constrained uniquely.
Instead, the same best-fit solution in figure \ref{GL} can be reproduced by multiple sets of ($M$, $R$, $D$, $B$).
For example, a set of (2.02$M_{\odot}$, 14.1 km, 0.81 kpc, $6\times 10^{13}$ G) and a set of (2.05$M_{\odot}$, 9.5 km, 0.81 kpc, $23\times 10^{13}$ G) give essentially the same prediction, and hence the same $\chi^2$.

\begin{figure}
 \begin{center}
  \includegraphics[width=0.5\textwidth]{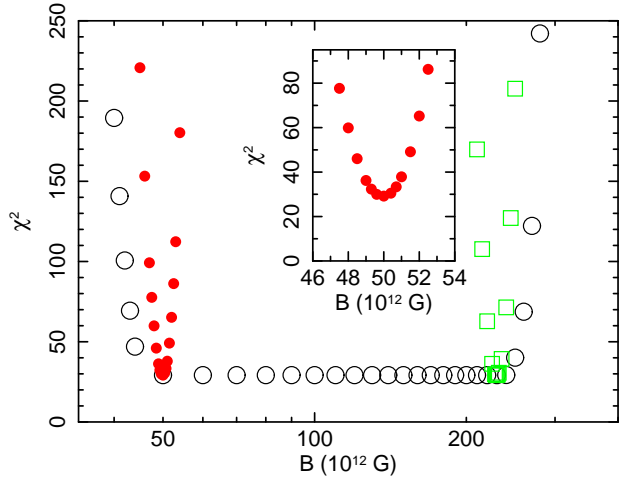}
 \end{center}
 \caption{The $\chi^2$ value of the fit to figure \ref{GL} with equation (1), shown as a function of the assumed $B$.
 Open black circles show the results when $M$, $R$, and $D$ are allowed to take free values within the allowed ranges.
Red points exemplify the case with $M=1.90M_{\odot}$, $R=14.6$ km, and $D=0.77$ kpc, whereas green ones assume $M=2.05M_{\odot}$, $R=9.5$ km, and $D=0.81$ kpc.
The inset shows a detail at $B=(46-54)\times10^{12}$ G.}
\label{bchi}
\end{figure}

Black circles of figure \ref{bchi} represents the minimum $\chi^2$ values as a function of the assumed $B$, where $M$, $R$, and $D$ are allowed to more freely within their respective constraints.
We thus find that the same best-fit solution is obtained over a range of $B=(5-23) \times 10^{13}$ G.
Below $B=5\times 10^{13}$ G, the fit $\chi^2$ starts increasing because $R$ is saturated at $R=15$ km.
In the same way, the fit worsening above $B=23\times 10^{13}$ G is due to $R$ fitting the assumed floor at $R=9.5$ km.
This is mainly because the value of $\mu \sim \frac{1}{2} {BR^3}$ is roughly fixed by the zero-cross point in figure \ref{GL}, that is, the spin up/down boundary \citep{Takagi2016}.
The effect of changing $B$, with the other parameters kept fixed, is shown by two blue lines in figure \ref{GL}.
Under this constraint, only the zero-cross, only the zero-cross point moves with the slope almost unchanged, the data favors $B=8\times 10^{13}$ G.

Red and green points in figure \ref{bchi} represent the behavior of $\chi^2$ when the set of ($M$, $R$, $D$) are fixed to the best values at $B=5\times 10^{13}$ G and those for $B=23\times 10^{13}$ G, respectively.
When $M$ and $D$ are fixed, the increase of $\chi^2$ thus becomes steeper than the case when they are free.
In short, the present study indicates $B=(5-23)\times 10^{13}$ G even allowing $M$, $R$, and $D$ to vary freely over the assumed uncertainty ranges.

\begin{figure}
 \begin{center}
  \includegraphics[width=0.5\textwidth]{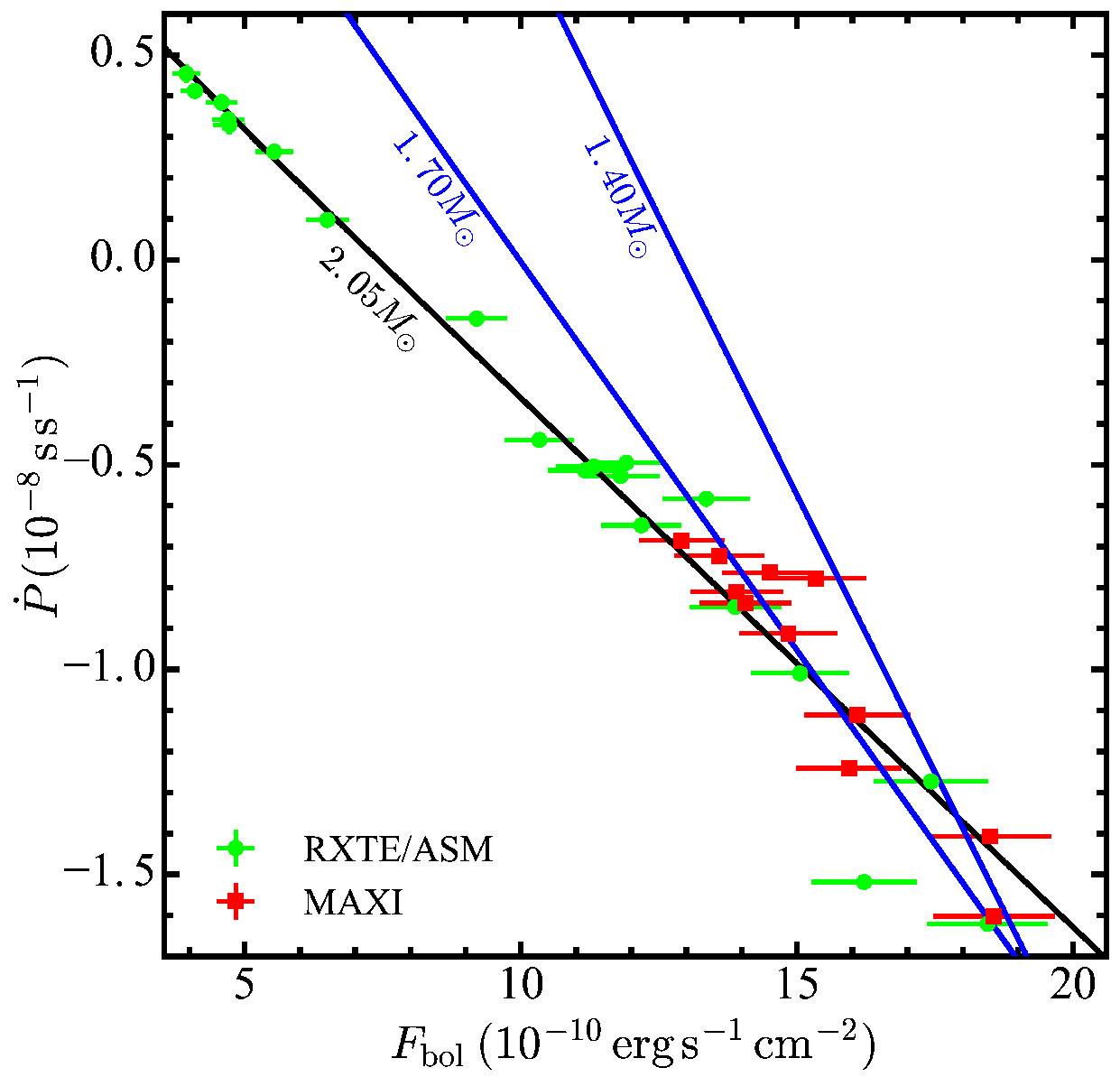}
 \end{center}
\caption{The data points and the black line are the same as in Fig.\ref{GL}. The blue lines are for $M=1.70M_{\odot}$ and $M=1.40M_{\odot}$ with the same $R$, $D$, and $B$ as in the black line.}
\label{GLM}
\end{figure}

As represented by equation (9) in \citet{Takagi2016}, the slope of the GL79 relation depends mainly on $M$ and $D$.
Since $\dot{P}$ is proportional to $I^{-1}$, and hence to $M^{-1}$, the slope flatters as $M$ increases.
Figure \ref{GLM} presents the same dataset as figure \ref{GL}, but is meant to show how the GL79 prediction depends on $M$.
Thus, the data prefer a relatively high value of $M$.

So far, we have confirmed the argument by \citet{Takagi2016} that figure \ref{GL} can constrain, with its zero-cross point and the data slope, two out of the four model parameters ($M$, $R$, $D$, and $B$).
In other words, specifying two of them can allow us to estimate the remaining two through the GL79 fit to figure \ref{GL}.
Choosing $D$ and $B$ as the input two parameters, figure \ref{massradi} shows how $M$ and $R$ are determined.
The red and blue lines in figure \ref{massradi} represent grids of constant $D$ and $B$, respectively.
Thanks to the accurate distance determination, $M$ is relatively well constrained, although we need to consider systematic errors involved in equation (1).
In these arguments using figure \ref{massradi}, we assumed that the radius is in the range of $R=9.5-15$ km, based on a recent equation of state of NSs \citep{Ozel2016} and the GW170817 observations \citep{Bauswein2017}.

\subsection{Systematic model uncertainties}
In the previous section,
we considered how the data can constrain  $M$, $R$, $D$, and $B$, 
when the first three of them are allowed to move freely 
over their pre-defined uncertainty ranges, 
whereas $B$ can take any positive value. 
However, we have yet to considered another source of systematic error, 
namely, the uncertainty in the GL79 model itself.
This may be expressed by multiplying the right side of 
equation (1) by a normalization factor $A$, 
and allowing it to take values rather than unity.

Along the above line of argument, \citet{Takagi2016} showed
that the MAXI/GSC data of the low-mass X-ray pulsar 4U~1626--67
can be explained very well with $A \approx 1.0$.
Analyzing the MAXI/GSC and Fermi/GBM data of 12 Be pulsars
in a similar manner, \citet{Sugizaki2017} argued
that the values of $A$ also has a mean around unity,
but it scatters from $\sim 0.3$ to $\sim 3$ among the 12 sample objects.
Furthermore, \citet{Sugizaki2017} successfully decomposed 
this scatter into the following three major sources.
One comes from the model assumption 
that the emission of a pulsar is completely isotropic,
which is obviously not warranted because X~Persei show strong pulsations.
As adopted by \citet{Sugizaki2017},
this uncertainty may amount to a factor of 2, after \citet{Basko1975}.
Another uncertainty lies is the angle between the magnetic axis and the accretion plane;
it can affect the dipole magnetic-field strength at large distances
by another factor of 2 \citep{Wang1997}. 
The last uncertainty is that in the distance.

\begin{figure}
 \begin{center}
  \includegraphics[width=0.5\textwidth]{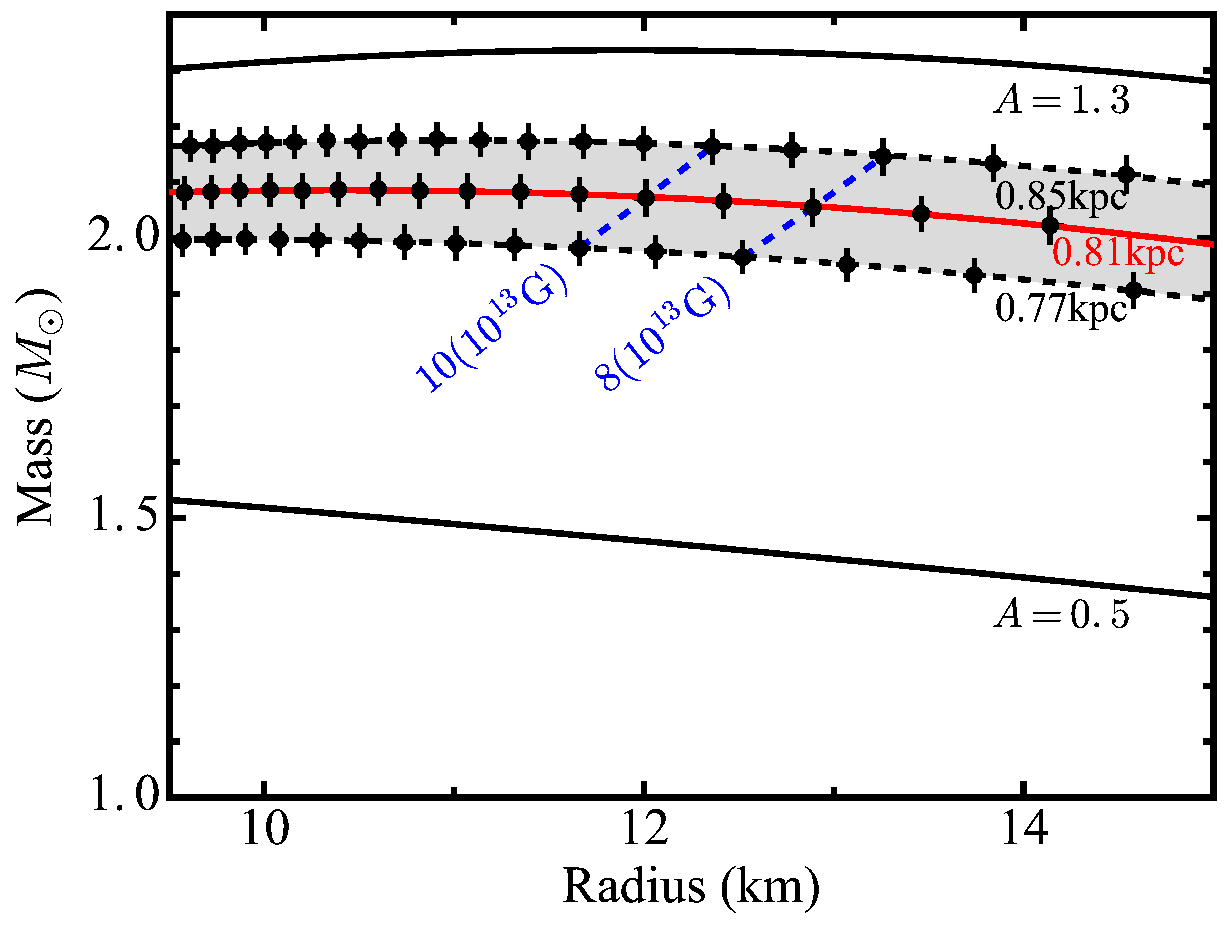}
 \end{center}
\caption{Sets of ($M$, $R$) accepted by the present data.
The constant-$B$ and constant-$D$ grids are represented by dashed blue lines and dashed black lines, respectively.
In particular, the $M-R$ relation fixing $D=0.81$ kpc is represented by the red line.
The solid black lines are those when the normalization of equation (1) is chosen as $A=0.5$ and $A=1.3$, both with $D=0.81$ kpc.}
\label{massradi}
\end{figure}

In the present case, the last uncertainty, namely that in $D$, can be ignored, 
because it is rather small, and was already considered before introducing $A$.
Therefore, we may presume that $A$ takes a value from 0.5 to 1.9.
Then, the model fitting was repeated by multiplying $A$ onto right side of equation (1),
and changing it from 0.5 to 1.9.
For example, two lines with $A = 0.5$  and $A = 1.3$ are shown in figure \ref{massradi},
where $D= 0.81$ kpc is fixed.
As a result, the mass and magnetic-field ranges have
somewhat increased to $M =1.3-2.4~M_\odot$
and $B = (4-25)\times 10^{13}$ G, respectively.

\section{Discussion and Conclusion}
Using the RXTE/ASM and MAXI/GSC data spanning altogether 22 yr,
we studied the spin-period change rate $\dot{P}$ 
and the bolometric luminosity $L$ of X~Persei.
Then, $\dot{P}$ has been confirmed to behave 
as a single-valued and approximately linear function of $L$ (figure \ref{GL}),
covering both the spin-down ($\dot{P}>$0) phase before MJD 52200
and the spin-up ($\dot{P}<0$) phase afterwards.
At the present spin period of $P \sim 837$ s,
the torque equilibrium condition ($\dot{P}=0$; the zero-cross point in figure \ref{GL}) 
is realized at  $L=5\times10^{34}$ erg s$^{-1}$ assuming $D=0.81$ kpc.
In addition, the orbital period of X~Persei
has been updated to $251.0^{+0.2}_{-0.1}$ d.

\subsection{The magnetic field of the neutron star in X~Persei}
By fitting the observed $\dot{P}$ vs $L$ relation
with the theoretical GL79 model,
incorporating its observational calibrations \citep{Takagi2016, Sugizaki2017},
we have estimated the surface dipole magnetic field of X~Persei 
as $B=(4-25) \times 10^{13}$ G.
Since the equilibrium values of $L$ and $P$ together specify 
the magnetic dipole moment ($\propto BR^3$) rather than $B$ itself,
the derived estimate of $B$ is most strongly 
affected by the uncertainty in $R$.
In fact, as in figure \ref{bchi},
the lower and upper limits of the derived $B$ 
respectively correspond to the upper bound of $R=15$ km
and the lower bound of $R=9.5$ km,
which we assumed as conservative limits.
If $R$ is further restricted in future studies,
e.g., by gravitational wave observations,
the range of $B$ will become narrower accordingly.
In any way, the present results show
that X~Persei has a significantly stronger magnetic field 
than ordinary accretion-powered pulsars ($B\sim10^{12}$ G).
\citet{Klus2014} investigated the relation between $\dot{P}$ and $L$ of 42 Be/X-ray binaries and showed that,
if those systems are in near spin equilibrium,
more than half of them have magnetic fields over the quantum critical level of $4.4 \times 10^{13}$ G.
They estimated the magnetic field of X~Persei would be $4.2 \times 10^{13}$ G,
which is consistent with our results.

Our result is consistent with the suggestions from 
spectral studies by \citet{Doroshenko2012} and \citet{Sasano2015}.
These authors noticed that the spectrum of X~Persei,
extending to 160 keV \citep{Lutovinov2012} without a clear cutoff,
is qualitatively distinct from those of ordinary X-ray pulsars,
which decline steeply above 20--30 keV
at least partially due to the presence of CRSFs \citep{Makishima2016}.
Because the spectral cutoff of X~Persei
must appear at rather high energies,
its CRSF, if any, would be present above $\sim 400$ keV,
corresponding to $B\gtrsim 4 \times 10^{13}$ G.

As mentioned in section 1,
the spectrum of X~Persei exhibits
a shallow and broad dip at $\sim 30$ keV \citep{Coburn2001, Lutovinov2012}.
\citet{Coburn2001} interpreted it as a CRSF, 
and proposed a value of $B\sim 10^{12}\,\rm G$.
However, \citet{Doroshenko2012} fitted the spectrum of 
X~Persei successfully with two Comptonization models
(to be interpreted as thermal and bulk Comptonization),
without invoking a CRSF structure.
A very similar result was obtained by \citet{Sasano2015}
from a broad-band Suzaku spectrum of X~Persei.
Thus, the dip around 30 keV can be better interpreted as a 
"crossing point" of the two Comptonization components, 
rather than a CRSF.

\subsection{Luminosity of X~Persei}
The GL79 relation which we employed, namely equation (1), assumes mass accretion from an accretion disk.
This condition is probably satisfied in 4U~1626--67, and in the 12 Be pulsars studied by \citet{Sugizaki2017}.
Specifically, these Be binaries are considered to be accreting from the circumstellar envelopes of their Be-type primaries (e.g., \cite{Okazaki2013}),
through accretion disks that form near the pulsars.
Compared to these typical Be X-ray binaries, X~Persei has an unusually wide and nearly circular orbit with $a_{\rm x}\sim 2$ a.u. \citep{Delgado2001},
together with a very low luminosity that lacks orbital modulation.
In spite of these peculiarities, we believe that X~Persei is also fed with an accretion disk rather than via stellar-wind capture, for the following three reasons.

The first reason supporting the disk-accretion scheme
is that the primary star of X~Persei, of a spectral type (O9.5III-B0V)e \citep{Lyubi1997}, is not expected to launch massive stellar winds.
Its luminosity of $\sim 3 \times 10^{4}~L_\odot$
and the empirical stellar wind intensity (Cox et al. 2000)
predict a wind mass-loss rate of $\lesssim 10^{-7.5}~M_\odot$ yr$^{-1}$.
At a distance of 2 a.u. (appropriate for the NS of X~Persei)
from the star,
and assuming a typical wind velocity of $10^8$ cm s$^{-1}$,
this would yield a wind-capture X-ray luminosity of  $\lesssim 10^{32}$ erg s$^{-1}$,
which is much lower than the actually observed values of $10^{34-35}$ erg s$^{-1}$.
Second, the spectra of X~Persei show
neither strong iron-K emission lines (with an equivalent width
of at most 7.5 eV; \cite{Maitra2017}), nor variable high absorption.
According to \citet{Makishima1990},
these properties of X~Persei supports the scheme of mass accretion 
from the Be envelope, rather than from the stellar winds.
Finally, the Be envelope is likely to be actually feeding the pulsar,
because a positive correlation was observed
between the optical H$\alpha$ line intensity and the X-ray flux
over 2009--2018 \citet{Zamanov2018},
involving the possible 7-yr periodicity,
although no clear correlation was observed before 2009 \citep{Smith1999, Reig2016}.

From the above arguments,
X~Persei is considered to have the same accretion scheme as the other Be X-ray binaries.
Its unusually low luminosity, without orbital modulation,
can be attributed to to its very wide and nearly circular orbit,
along which the stellar envelope is expected to have a rather low density.
Incidentally, the possible 7-yr X-ray periodicity (Section 2.4),
correlated with the H$\alpha$ variations \citep{Zamanov2018},
may be produced by some long-term changes in the circumstellar envelope,
as pointed out by \citet{Laplace2017}.

\subsection{Mass of the neutron star in X~Persei}
While the zero-cross point of the $L$ vs $\dot{P}$ correlation 
in figure \ref{GL} is most sensitive to the magnetic dipole moment of the pulsar,
the  slope of the correlation specifies mainly $M$ \citep{Takagi2016}.
Thus, assuming $D=0.81\pm0.04$ kpc, 
and neglecting  the systematic model uncertainty (Section 4.2),
the NS in X~Persei is constrained 
to have $M=2.03\pm0.17M_{\odot}$ as in figure \ref{massradi}.
Therefore, the object is suggested to be somewhat more massive 
than typical NSs with $M\sim 1.4M_{\odot}$.

As shown in figure \ref{massradi}, 
the systematic GL79 model uncertainty affects the estimate of $M$,
and its inclusion widens the range as  $M = 1.3-2.4~M_{\odot}$.
Then, the canonical value of $M=1.4~M_\odot$ can no longer be excluded.
Nevertheless, in order for the object to have $M=1.4~M_\odot$,
we need to invoke the smallest value of $A \sim 0.5$.
This would in turn require, for example, 
that the emission from X~Persei is strongly beamed away from our line of sight,
and we are underestimating its spherically averaged luminosity by a factor of 2.
Thus, the data still favor a relatively high mass (e.g., $\sim 2~M_\odot$).

Let us tentatively assume that the NS in X~Persei is indeed massive.
Then, this must be from its birth,
because  the accretion rate of X~Persei 
($\sim 1 \times 10^{-12}~M_\odot$ yr$^{-1}$) and its estimated life time ($\sim 10^7$ yr)
mean a negligible mass increase due to accretion.
We hence arrive at an interesting possibility,
that a fraction of NSs could be born with 
a strong magnetic field ($B\geq10^{13}$ G) together with a rather high mass.
Such a native difference, if true, might in turn 
reflect different mechanisms of the NS formation.
Even excluding the case of accretion-induced collapse of white dwarfs,
NSs in Be X-ray binaries may be produced 
via supernovae of either electron-capture type, 
or gravitational core-collapse type.
As discussed by \citet{Kitaura2006},
the former types are likely to leave rather standard NSs with $\sim1.4M_{\odot}$ and $B\sim10^{12}$ G,
as is possibly the case with the Crab pulsar \citep{Moriya2014}.
In contrast, core-collapse supernovae might sometimes yield 
NSs with a higher $M$ and a stronger $B$,
as represented by X~Persei.
Even in that case, it is an open question 
whether a higher mass necessarily lead to a stronger field,
or these two quantities scatter independently.

\section*{Acknowledgement}
The authors are grateful to all MAXI team members for scientific analysis.

%
%
%
%
%
%
%
%
%
%
%



\end{document}